\documentclass[conference]{IEEEtran}
\IEEEoverridecommandlockouts
\usepackage{cite}
\usepackage{amsmath,amssymb,amsfonts}
\usepackage{algorithmic}
\usepackage{graphicx}
\usepackage{textcomp}
\usepackage{xcolor}
\usepackage{hyperref}
\def\BibTeX{{\rm B\kern-.05em{\sc i\kern-.025em b}\kern-.08em
    T\kern-.1667em\lower.7ex\hbox{E}\kern-.125emX}}

\begin{document}

\title{Advances in Privacy Preserving Federated Learning to Realize a Truly Learning Healthcare System
}

\author{\IEEEauthorblockN{Ravi Madduri}
\IEEEauthorblockA{\textit{Data Science and Learning Division} \\
\textit{Argonne National Laboratory}\\
Lemont, IL USA  \\
madduri@anl.gov}
\and
\IEEEauthorblockN{Zilinghan Li}
\IEEEauthorblockA{\textit{Data Science and Learning Division} \\
\textit{Argonne National Laboratory}\\
Lemont, IL USA  \\
zilinghan.li@anl.gov}
\and
\IEEEauthorblockN{Tarak Nandi}
\IEEEauthorblockA{\textit{Data Science and Learning Division} \\
\textit{Argonne National Laboratory}\\
Lemont, IL USA  \\
tnandi@anl.gov}
\and
\IEEEauthorblockN{Kibaek Kim}
\IEEEauthorblockA{\textit{Mathematics and Computer Science} \\
\textit{Argonne National Laboratory}\\
Lemont, IL USA  \\
kimk@anl.gov}
\and
\IEEEauthorblockN{Minseok Ryu}
\IEEEauthorblockA{ \textit{School of Computing and Augmented Intelligence} \\
\textit{Arizona State University}\\
Tempe, AZ USA \\
minseok.ryu@asu.edu}
\and
\IEEEauthorblockN{Alex Rodriguez}
\IEEEauthorblockA{\textit{Data Science and Learning} \\
\textit{Argonne National Laboratory}\\
Lemont, IL USA  \\
a.rodriguez@anl.gov}
}

\maketitle

\begin{abstract}
    The concept of a learning healthcare system (LHS) envisions a self-improving network where multimodal data from patient care are continuously analyzed to enhance future healthcare outcomes. However, realizing this vision faces significant challenges in data sharing and privacy protection. Privacy-Preserving Federated Learning (PPFL) is a transformative and promising approach that has the potential to address these challenges by enabling collaborative learning from decentralized data while safeguarding patient privacy. This paper proposes a vision for integrating PPFL into the healthcare ecosystem to achieve a truly LHS as defined by the Institute of Medicine (IOM) Roundtable.
\end{abstract}


\section{Introduction}
\begin{figure*}[h]
\centerline{\includegraphics[width=\linewidth]{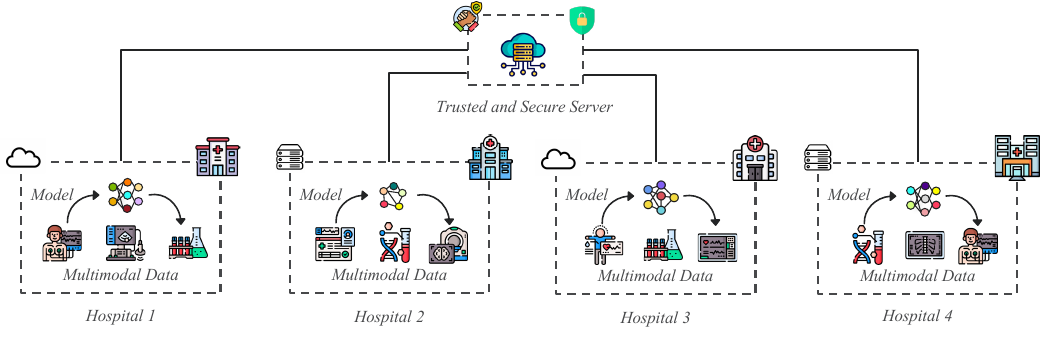}}
\caption{An envisioned privacy-preserving federated learning framework for a truly learning healthcare system: Under the coordination of a trusted and secure server, multiple hospitals collaboratively train robust, generalized machine learning models using multimodal biomedical data stored in their cloud or on-premise facilities. With continuous learning capabilities integrated into the framework, the models can detect and avoid  performance degradation, adapt dynamically in real time to any shifts in data distributions,  availability of new patient data, and evolving health trends.}
\label{fig:overview}
\end{figure*}
Biomedical health data is often generated across various instruments, hospitals, or departments that are administratively and geographically disparate. Efforts are usually required to move data to a central location for analysis, requiring considerable time and infrastructure investments to facilitate. These efforts result in the creation of clinical data warehouses and disease-specific consortiums, which often evolve into data silos making data discovery and sharing cumbersome. To realize a truly learning healthcare system as defined by the Institute of Medicine—and to deliver personalized prognosis, diagnosis, and treatment planning—it is essential to learn from and synthesize multimodal biomedical data collected and stored across diverse administrative and geographical boundaries. Biomedical data are inherently multimodal, encompassing electronic health records, genomic information, medical imaging, laboratory results, and data from wearable devices, among others. However, these rich datasets are often soloed due to administrative disparities, privacy concerns, and regulatory constraints, which impede comprehensive analysis and integration.
Furthermore, additional efforts are needed to extract, transform, and load data to gain insights. Recent advances in artificial intelligence (AI) have demonstrated the potential for rapid insight generation from scientific and biomedical data \cite{gulshan2016development, esteva2017dermatologist, miotto2018deep}, however, developing effective biomedical AI models requires substantial data and computing resources \cite{kaplan2020scaling}. The conventional paradigm of building large-scale AI models includes collecting extensive data and training models at a central location. With the increasing volume and velocity of data generation, such a centralized model development paradigm is becoming impractical in some scientific domains. In biomedical health, the central collection of data from multiple data generation sources, especially those across different administrative boundaries, is often not possible due to data privacy and federal policies like HIPAA\cite{HIPAA1996}, GDPR{\cite{GDPR2016}}. Nonetheless, models trained on limited data sets from a single source frequently fail to perform in real-world situations (performance degradation) when the distribution of real-world data differs from the distribution of the training data, a phenomenon known as model shift.
\section{Privacy Preserving Federated Learning}
Several approaches have been proposed to alleviate model drift, detect performance degradation and facilitate the development of robust AI models that perform reliably in the real world.  Federated learning (FL), a distributed learning approach where a global model is created by aggregating model weights from models trained on data at various sites, can address the challenges of building robust AI models that are resistant to model drift without direct data sharing \cite{konevcny2016federated,mcmahan2017communication, flconcept, FLadvances}. Though no data are directly shared among sites, however, FL by itself does not guarantee the privacy of data, because the information extracted from the communication of FL algorithms can be accumulated and utilized to infer the private local data used for training \cite{zhao2020idlg, geiping2020inverting}. Differential Privacy (DP)\cite{executive_order_2023}, a privacy-enhancing technology (PET),  when integrated with FL is shown to prevent data reconstruction by attackers. DP adds noise to model updates to prevent accurate data reconstruction by attackers. Infrastructure challenges in the implementation of FL at scale include dealing with heterogeneity of computational resources available to researchers, identity and access management challenges to setup end-to-end secure federations, ease of use in setting up and running FL experiments, addressing privacy issues for different data types, and FAIRness constructs when performing AI experiments to enable reproducibility \cite{ravi2022fair, huerta2023fair}.

To address these issues, we have developed the Advanced Privacy-Preserving Federated Learning (APPFL) \cite{ryu2022appfl,li2024advancesappflcomprehensiveextensible} framework with advances in differential privacy \cite{dwork2006differential}. APPFL enables the training of AI models in a distributed setting across multiple institutions, where sensitive data are located, with the ability to scale on distributed, heterogeneous computing resources to help create robust, trust-worthy AI models in biomedical health applications where data privacy is essential. Setting up a secure FL experiment across administrative boundaries that involves heterogeneous high-performance computational resources across distributed sites or cloud computing facilities  requires technical capabilities that may not be available for all. Additionally, most existing PPFL frameworks typically involve downloading and configuring complex software, manually creating trust boundaries to enable secure gradient aggregation, and understanding the technical details of underlying deep learning software stack \cite{beutel2020flower, roth2022nvidia}, all of which can be cumbersome and technically demanding. Therefore, to reduce these barriers and empower domain experts to leverage PPFL, we created the Advanced Privacy-Preserving Federated Learning as a service (APPFLx) \cite{li2023appflx} platform, which streamlines cross-silo PPFL using an easy-to-use web interface for managing, deploying, analyzing, and visualizing PPFL experiments. APPFLx ensures secure federations using end-to-end strong Identity and Access Management via Globus Auth \cite{tuecke2016globus}, enabling members to create new federations or join ones using their institutional identities, perform privacy-preserving training on datasets at their respective institutions, and securely share the model weights with the service for secure aggregation. Additionally, determining quality of training data is paramount in developing biomedical health AI models as low-quality or biased data leads to ineffective and unreliable AI models. Therefore, to ensure the integrity of training data before committing significant computing resources to training jobs, APPFL incorporates the AI Data Readiness Inspector (AIDRIN) \cite{hiniduma2024ai}, an open-source toolkit. AIDRIN integration allows for a distributed quantitative assessment of data readiness, providing data scientists with vital metrics that streamline data preparation and facilitate informed decisions regarding the suitability of data for AI applications. It not only saves time but also optimizes the effort invested in the initial stages of model development.

While APPFL and APPFLx have significantly streamlined traditional PPFL experiments and facilitated the training of unimodal biomedical AI models \cite{hoang2023enabling, li2024advancesappflcomprehensiveextensible}, there remains considerable progress to be made. As illustrated in Figure~\ref{fig:overview}, our envisioned PPFL framework aims to advance a truly learning healthcare system that is capable of providing more precise diagnoses, prognoses, treatments, and preventative measures. Such a framework utilizes a trusted server, secured with privacy-enhancing techniques and robust cybersecurity algorithms, to orchestrate PPFL experiments among client hospitals with heterogeneous cloud or on-premise computing and storage systems. Through this collaboration, multiple healthcare delivery organizations can jointly train robust machine learning models that effectively generalize across diverse patient populations by leveraging multimodal biomedical data stored within their private infrastructures. Moreover, addition of continuous learning to PPFL framework, allows for early detection of any performance degradation, real-time adaptation of models to new patient data, and evolving health trends across multiple healthcare provides. In the following sections, we will provide details about the essential building blocks of this envisioned framework, including federated training for multimodal biomedical models, hierarchical FL for collaborative training beyond limited trust boundaries, federated continuous training, and cost-aware FL on the cloud.

\section{Federated Training of Multimodal Biomedical Models}
\begin{figure*}[h]
\centerline{\includegraphics[width=0.8\linewidth]{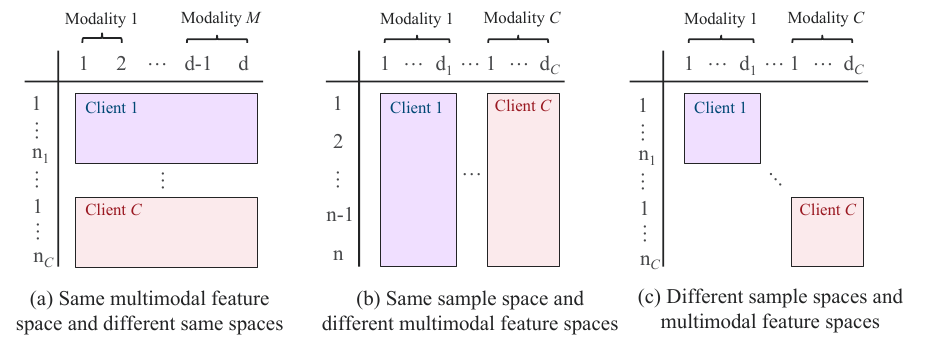}}
\caption{Different sample and multimodal feature distribution patterns among clients in multimodal federated learning. }
\label{fig:modalities}
\end{figure*}
Biomedical health data are inherently multimodal, encompassing electronic health records, genomic information, medical imaging, laboratory results, and data from wearable devices, among others \cite{stahlschmidt2022multimodal, tu2024towards}. In practice, clinicians typically integrate multiple available data types when making diagnoses or treatment decisions, most current biomedical health AI models are limited to specific tasks based on single data sources, such as imaging or text \cite{acosta2022multimodal}. Multimodal AI models, which can utilize data from diverse sources like genetics, imaging, clinical records, and environmental factors, offer a promising solution to this limitation. These models have the potential to handle the complexity and high dimensionality of biomedical data, which is crucial for understanding interactions within biological systems and predicting complex diseases. As biomedical data increasingly become multimodal, such models are poised to revolutionize personalized medicine, digital clinical trials, and real-time health surveillance \cite{acosta2022multimodal}. Currently, most FL applications in biomedicine use traditional horizontal FL, where all the clients share the same feature space and collaboratively train the same model architecture under the orchestration of a central server. These applications primarily rely on uni-modal data such as electronic health records and medical imaging \cite{fl_cancer,fl_medimg,ogier2022flamby,hoang2023enabling}. However, given the inherently multimodal nature of biomedical data, it is important to explore the feasibility of employing PPFL to train robust, generalized multimodal AI models by using the diverse biomedical health data available across diverse administrative and geographical boundaries.

\subsection{Multimodal Learning}
Some common modalities in biomedicine include images (e.g., MRI, histology), tabular data (e.g., gene expression), and text (e.g., clinical data). Multimodal learning aims to utilize these different data modalities to provide the most informative predictions. Combining data from multiple sources enhances the information available beyond what any individual modality can provide, and integrating weak signals across modalities can help overcome noise present in a single modality.

For example, integrating molecular, imaging, and clinical data can improve the quality and accuracy of biomarkers for cancer, offering a more comprehensive understanding of the disease, instead of any of these modalities individually. Single-modality data, such as radiology scans or gene expression data from bulk RNASeq, are often found to be insufficient for capturing the complex heterogeneity of cancer. While radiology scans provide macroscopic information with spatial context, bulk RNASeq gene expression data offers molecular-level insights without the spatial context. By integrating these orthogonal data sources, we can capture complementary aspects of tumor biology.

Different data modalities—such as clinical notes, medical images, genomic data, and sensor readings—often contain complementary information and may vary in quality. When these diverse modalities are effectively integrated, they can provide a more comprehensive and informative view than any single data type alone, leading to more accurate prognoses, diagnoses, and treatment plans. However, directly combining such heterogeneous data is often impractical due to differences in data formats, structures, scales, and the complexities involved in processing them together.

To address this challenge, most multimodal models employ embeddings, transforming data from each modality into unified representations within a shared feature space. This approach facilitates the integration of different data types by encoding their essential features in a way that makes them compatible for joint analysis. The method of embedding generation is critical, as it significantly affects the level of inter-modal learning and, consequently, the overall performance of the model. Effective embedding strategies enhance the model's ability to learn from the combined data, leading to more robust and insightful outcomes in biomedical research and healthcare applications.

Multimodal AI models generate joint representations of heterogeneous data modalities through data fusion, which can be broadly categorized into three types: early fusion, joint fusion, and late fusion \cite{steyaert2023multimodal}. In early fusion, the embeddings from the different modalities are generated independently and then combined to act as input to a downstream task-specific prediction model. In joint fusion, the embeddings from all the modalities are dynamically updated during training, allowing them to learn directly from the task objective as well as from the other modalities. In late fusion, instead of combining the embeddings from different modalities, the predictions from separate uni-modal submodels are combined into a final prediction. While joint fusion is theoretically the most informed way to learn the embeddings, it can be computationally expensive as it requires training the embedding generation models and the final prediction model simultaneously. 

Integrating data from multiple sources and modalities to train a single AI model can be challenging due to varying information content and quality across modalities and sources. The embeddings must be context-specific and capable of capturing relevant features from their respective modalities. They should have sufficient dimensionality to represent the underlying characteristics of the modalities that are relevant for the task, but not so high as to become computationally expensive for model training.  Such embedding generation methods can differ significantly among modalities, and are often strongly tied to the biological insights of the data they represent. For example, in histology whole slide images (WSI) or MRI scans, convolutional neural networks (CNNs) or vision transformer models can be used to generate embeddings after learning important features from a large number of training samples \cite{kang2023benchmarking}. For bulk RNASeq data available in tabular form, variational autoencoders (VAEs) can be used to generate lower dimensional context-specific embeddings from the very high dimensional gene expression data \cite{carrillo2024generation}. By carefully selecting embedding generation techniques tailored to each modality, it is possible to harmonize diverse data sources leading to improved model performance and more informative predictions.

\subsection{Integration of Multimodal Learning in a Federated Setting}
When training a multimodal model in a federated setting, there are three scenarios based on the distribution of the sample and multimodal feature spaces. In the first scenario, when all clients shared the same multimodal feature space, as depicted in Figure~\ref{fig:modalities}-a, a multimodal model can be trained directly using traditional FL algorithms, where each client trains an individual model with the same architecture and shares the locally trained model parameters for aggregation \cite{xiong2022unified, che2023multimodal}. In the second scenario, the clients share the same sample space but possess different data modalities, as shown in Figure~\ref{fig:modalities}-b, making vertical FL (VFL) an appropriate solution. In VFL, each client trains a local model on its respective data modalities, generates latent representations, and sends these to the server to update the global model. The server then sends back gradients for the corresponding latent representations, allowing each client to update its local model \cite{fu2021vf2boost, liu2024vertical}. However, VFL typically requires data identifiers to align distributed samples, which may not always be feasible in biomedical applications. In the third scenario, where clients have different sample and multimodal feature spaces and lack unique identifiers for aligning samples (Figure~\ref{fig:modalities}-c), training a multimodal model collaboratively becomes more challenging. One potential solution is for each client to train a sub-model on its local data modalities and share it with other clients to collectively build a larger multimodal model using intermediate or late fusion techniques. Another approach involves having each client train a local model on its private data modalities and transmit the latent representations of publicly available multimodal data to the server for sample alignment \cite{yu2023multimodal}.  Additionally, generative models could be used to create synthetic data that mimics the distribution of each client's multimodal data, enabling alignment without compromising sensitive information. While promising, these approaches remain an open area of research, requiring further investigation to address the complexities of multimodal FL.

\section{Hierarchical FL for Collaborative Training Beyond the Trust Boundaries}

In FL, while no data are directly exchanged between the server and clients, private and confidential training data can still be vulnerable to reconstruction from shared model gradients through gradient inversion techniques \cite{zhao2020idlg, geiping2020inverting, hatamizadeh2023gradient, hoang2023enabling}. As a result, some clients may be reluctant to share trained model parameters or gradients with an untrusted central server. For instance, small clinics within a geographic region might prefer to share model parameters exclusively with a trusted or affiliated local hospital. However, this limited trust boundary could result in biased localized datasets, restricting the model's ability to generalize across diverse populations. Additionally, the need for diverse data becomes even more pressing in biomedical health applications involving rare diseases, where datasets are often small and fragmented across multiple institutions. Limiting data sharing with a single trusted entity may reduce the ability to capture complex patterns that are essential for understanding and diagnosing such conditions.

\begin{figure}[h]
\centerline{\includegraphics[width=\linewidth]{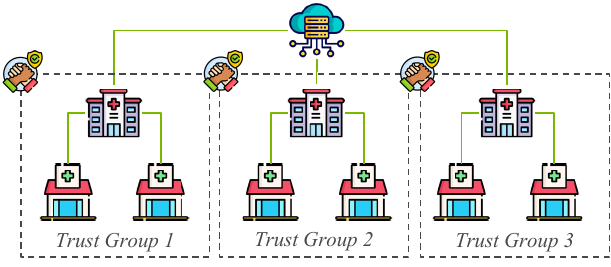}}
\caption{Hierarchical federated learning helps to connect several small trust group to a larger federation.}
\label{fig:hfl}
\end{figure}

Hierarchical federated learning (HierFL) \cite{abad2020hierarchical, briggs2020federated} offers a solution to this challenge by enabling broader participation while maintaining trust. As illustrated in Figure~\ref{fig:hfl}, HierFL allows groups of clients, such as small clinics, to first share their local model parameters with an intermediate server trusted by the clients, such as a large local hospital. This intermediate server aggregates the local parameters, obscuring individual client information. Aggregated models from multiple intermediate servers, representing different trusted groups, are then sent to a central root server for further global aggregation. The resulting global model, now trained on a more diverse and representative dataset, is subsequently distributed back to the clients for further local training. HierFL enhances privacy by introducing an additional layer of aggregation with trusted intermediate servers, thus encouraging wider participation by clients beyond their immediate trust boundaries. This intermediate aggregation process safeguards individual model details while enabling the creation of a global model that generalizes better across varied populations, making it particularly valuable for biomedical health applications.

\section{Federated Continuous Learning for Multimodal Biomedical Models}

Continuous learning, the process by which models are continuously updated and improved as new data becomes available \cite{thrun1995lifelong,parisi2019continual}, is essential in biomedical applications \cite{lee2020clinical}. It enables healthcare systems to remain adaptive and responsive, evolving to meet the ever-changing needs of patients across different settings. In a field where medical knowledge, treatments, and patient conditions are constantly evolving, continuous learning ensures that models stay relevant and resistant to performance degradation by incorporating real-time data. By continuously integrating new information from sources like patient records, clinical trials, and wearable devices, healthcare models can provide more accurate and robust predictions, diagnostics, and treatment recommendations. This dynamic approach allows healthcare systems to offer more personalized, timely care while improving outcomes for diverse patient populations. Ultimately, continuous learning helps reduce healthcare disparities and creates a more effective, data-driven healthcare ecosystem that can evolve in response to emerging challenges, thus enabling true learning healthcare systems (LHS).

Integrating continuous learning capabilities into a PPFL framework offers the framework the ability to timely adapt to new data from diverse medical institutions while maintaining data privacy. This integration allows models to stay up-to-date with the latest medical advancements, patient information, and social health trends, leading to improved predictive accuracy treatment recommendations. Additionally, continuous learning within a PPFL framework ensures that the models evolve to reflect the diversity of healthcare settings and populations, reducing biases and enabling better generalization. Ultimately, this combination enhances the capacity of healthcare systems to provide real-time, data-driven insights without compromising patient privacy. Additional capabilities need to be developed to calculate the optimal privacy budget for biomedical multimodal datasets to prevent model inversion attacks but allow for improved model performance.

\begin{figure}[h]
\centerline{\includegraphics[width=\linewidth]{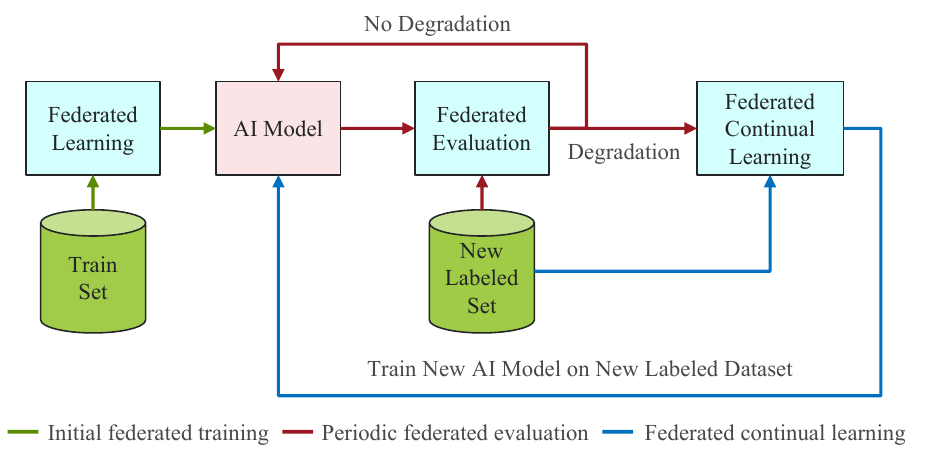}}
\caption{Federated continuous learning workflow with a federated evaluation feedback loop for timely performance degradation detection.}
\label{fig:continual}
\end{figure}

Additionally, the PPFL framework can be extended with federated evaluation and monitoring capabilities to enable timely assessment and monitoring of the model’s performance across diverse and distributed client populations. As shown in Figure~\ref{fig:continual}, whenever a biomedical AI model is trained via FL, the system will periodically and systematically evaluate its key performance metrics such as accuracy, precision, and recall in a federated setting to ensure the model remains effective under varying conditions and data distributions. Should the system detect any significant performance degradation, such as a decline in accuracy or a shift in data patterns, these insights can trigger an immediate, automated response via initiating a federated continuous learning process among the distributed clients to allow the model to adapt and retrain on new or evolving datasets without compromising data privacy. This dynamic feedback loop, when coupled with capabilities like AIDRIN, ensures that the model is regularly updated and optimized as needed, preventing performance degradation while addressing challenges such as data drift, population variance, or emerging trends in real-time. These capabilities significantly enhance the framework’s resilience and scalability, particularly in environments where data is continuously generated and needs to be rapidly incorporated into the model.

\section{Cost-Aware FL on the Cloud}

\begin{figure}[h]
\centerline{\includegraphics[width=\linewidth]{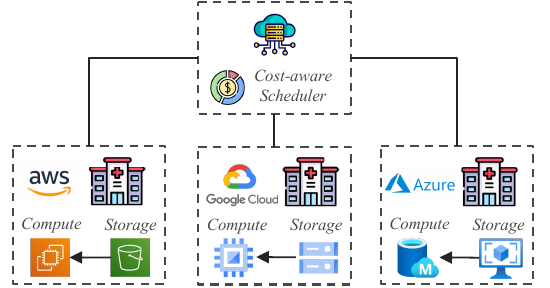}}
\caption{Using a server-side cost-aware scheduler to achieve cost-effective FL experiments among clients on the cloud.}
\label{fig:cloud}
\end{figure}

Training AI models is often expensive, especially when relying on GPU virtual machine instances from cloud providers like AWS, Google Cloud Platform, and Azure. As many medical institutes lack sufficient on-premise computing and storage resources, they are highly dependent on cloud providers for AI model training tasks. This creates a crucial need to make FL experiments more cost-effective, speeding up the transition of FL from experimental prototyping to real-world applications. Popular cloud providers offer a cost-saving option called spot instances, which allow users to bid for unused virtual machines at significantly lower prices than on-demand instances (usually 70\% to 90\% cheaper). However, as a compromise for the low price, these spot instances can be terminated at any time with a short notice when the cloud provider needs them for on-demand users. As a consequence, clients should checkpoint their local training status and notify the server when termination is imminent. On the server side, inspired by work like \cite{li2023fedcompass}, which uses a server-side computing-aware scheduler to enhance efficiency in heterogeneous environments, a cost-aware scheduler can be designed to make the FL experiments cost-effective, as shown in Figure~\ref{fig:cloud}. This scheduler would record the cost of training in real-time, manage the computing instances within a predefined training budget, reallocate resources for terminated instances, and perform aggregations even in the absence of offline clients. Implementing such cost-saving measures can significantly reduce financial barriers, making it more feasible to deploy FL more broadly within the healthcare ecosystems. 

\section{Conclusions}
In this paper, we propose that privacy-preserving federated learning (PPFL) could offer a viable pathway toward realizing a truly learning healthcare system by facilitating the training of robust multimodal biomedical models while preserving patient data privacy. Our envisioned PPFL framework goes beyond traditional limitations by incorporating supports for hierarchical FL, which extends trust boundaries and enhances collaborative potential among diverse healthcare institutions. It also integrates federated evaluation and continuous learning mechanisms that allow for real-time updates and improvements to models based on new data and insights, ensuring that the models remain relevant and accurate over time. Moreover, the envisioned framework includes the development of cost-effective algorithms that make the adoption of FL feasible even for institutions with limited resources. Through these innovations, PPFL can drive forward the evolution of healthcare towards more automated, responsive, and efficient practices.

% \begin{table}[htbp]
% \caption{Table Type Styles}
% \begin{center}
% \begin{tabular}{|c|c|c|c|}
% \hline
% \textbf{Table}&\multicolumn{3}{|c|}{\textbf{Table Column Head}} \\
% \cline{2-4} 
% \textbf{Head} & \textbf{\textit{Table column subhead}}& \textbf{\textit{Subhead}}& \textbf{\textit{Subhead}} \\
% \hline
% copy& More table copy$^{\mathrm{a}}$& &  \\
% \hline
% \multicolumn{4}{l}{$^{\mathrm{a}}$Sample of a Table footnote.}
% \end{tabular}
% \label{tab1}
% \end{center}
% \end{table}

% \begin{figure}[htbp]
% \centerline{\includegraphics{fig1.png}}
% \caption{Example of a figure caption.}
% \label{fig}
% \end{figure}

\section*{Acknowledgments}

This work was supported by the U.S. Department of Energy, Office of Science, Advanced Scientific Computing Research, under Contract DE-AC02-06CH11357.

\bibliographystyle{IEEEtran}
\bibliography{PPFL}

\begin{thebibliography}{10}
\providecommand{\url}[1]{#1}
\csname url@samestyle\endcsname
\providecommand{\newblock}{\relax}
\providecommand{\bibinfo}[2]{#2}
\providecommand{\BIBentrySTDinterwordspacing}{\spaceskip=0pt\relax}
\providecommand{\BIBentryALTinterwordstretchfactor}{4}
\providecommand{\BIBentryALTinterwordspacing}{\spaceskip=\fontdimen2\font plus
\BIBentryALTinterwordstretchfactor\fontdimen3\font minus \fontdimen4\font\relax}
\providecommand{\BIBforeignlanguage}[2]{{%
\expandafter\ifx\csname l@#1\endcsname\relax
\typeout{** WARNING: IEEEtran.bst: No hyphenation pattern has been}%
\typeout{** loaded for the language `#1'. Using the pattern for}%
\typeout{** the default language instead.}%
\else
\language=\csname l@#1\endcsname
\fi
#2}}
\providecommand{\BIBdecl}{\relax}
\BIBdecl

\bibitem{gulshan2016development}
V.~Gulshan, L.~Peng, M.~Coram, M.~C. Stumpe, D.~Wu, A.~Narayanaswamy, S.~Venugopalan, K.~Widner, T.~Madams, J.~Cuadros \emph{et~al.}, ``Development and validation of a deep learning algorithm for detection of diabetic retinopathy in retinal fundus photographs,'' \emph{jama}, vol. 316, no.~22, pp. 2402--2410, 2016.

\bibitem{esteva2017dermatologist}
A.~Esteva, B.~Kuprel, R.~A. Novoa, J.~Ko, S.~M. Swetter, H.~M. Blau, and S.~Thrun, ``Dermatologist-level classification of skin cancer with deep neural networks,'' \emph{nature}, vol. 542, no. 7639, pp. 115--118, 2017.

\bibitem{miotto2018deep}
R.~Miotto, F.~Wang, S.~Wang, X.~Jiang, and J.~T. Dudley, ``Deep learning for healthcare: review, opportunities and challenges,'' \emph{Briefings in bioinformatics}, vol.~19, no.~6, pp. 1236--1246, 2018.

\bibitem{kaplan2020scaling}
J.~Kaplan, S.~McCandlish, T.~Henighan, T.~B. Brown, B.~Chess, R.~Child, S.~Gray, A.~Radford, J.~Wu, and D.~Amodei, ``Scaling laws for neural language models,'' \emph{arXiv preprint arXiv:2001.08361}, 2020.

\bibitem{HIPAA1996}
{U.S. Congress}, ``{Health Insurance Portability and Accountability Act of 1996},'' \url{https://www.govinfo.gov/content/pkg/PLAW-104publ191/pdf/PLAW-104publ191.pdf}, 1996, public Law 104-191.

\bibitem{GDPR2016}
{European Parliament and Council of the European Union}, ``{Regulation (EU) 2016/679 of the European Parliament and of the Council of 27 April 2016},'' \url{https://eur-lex.europa.eu/eli/reg/2016/679/oj}, 2016, general Data Protection Regulation.

\bibitem{konevcny2016federated}
J.~Kone{\v{c}}n{\`y}, ``Federated learning: Strategies for improving communication efficiency,'' \emph{arXiv preprint arXiv:1610.05492}, 2016.

\bibitem{mcmahan2017communication}
B.~McMahan, E.~Moore, D.~Ramage, S.~Hampson, and B.~A. y~Arcas, ``Communication-efficient learning of deep networks from decentralized data,'' in \emph{Artificial intelligence and statistics}.\hskip 1em plus 0.5em minus 0.4em\relax PMLR, 2017, pp. 1273--1282.

\bibitem{flconcept}
Q.~Yang, Y.~Liu, T.~Chen, and Y.~Tong, ``Federated machine learning: Concept and applications,'' \emph{ACM Transactions on Intelligent Systems and Technology (TIST)}, vol.~10, no.~2, pp. 1--19, 2019.

\bibitem{FLadvances}
P.~Kairouz, H.~B. McMahan, B.~Avent, A.~Bellet, M.~Bennis, A.~N. Bhagoji, K.~Bonawitz, Z.~Charles, G.~Cormode, R.~Cummings \emph{et~al.}, ``Advances and open problems in federated learning,'' \emph{Foundations and Trends{\textregistered} in Machine Learning}, vol.~14, no. 1--2, pp. 1--210, 2021.

\bibitem{zhao2020idlg}
B.~Zhao, K.~R. Mopuri, and H.~Bilen, ``idlg: Improved deep leakage from gradients,'' \emph{arXiv preprint arXiv:2001.02610}, 2020.

\bibitem{geiping2020inverting}
J.~Geiping, H.~Bauermeister, H.~Dr{\"o}ge, and M.~Moeller, ``Inverting gradients-how easy is it to break privacy in federated learning?'' \emph{Advances in Neural Information Processing Systems}, vol.~33, pp. 16\,937--16\,947, 2020.

\bibitem{executive_order_2023}
\BIBentryALTinterwordspacing
{Executive Order No. 2023-24283}, ``Safe, secure, and trustworthy development and use of artificial intelligence,'' Federal Register, November 2023, accessed: 2023-11-01. [Online]. Available: \url{https://www.federalregister.gov/documents/2023/11/01/2023-24283/safe-secure-and-trustworthy-development-and-use-of-artificial-intelligence}
\BIBentrySTDinterwordspacing

\bibitem{ravi2022fair}
N.~Ravi, P.~Chaturvedi, E.~Huerta, Z.~Liu, R.~Chard, A.~Scourtas, K.~Schmidt, K.~Chard, B.~Blaiszik, and I.~Foster, ``Fair principles for ai models with a practical application for accelerated high energy diffraction microscopy,'' \emph{Scientific Data}, vol.~9, no.~1, p. 657, 2022.

\bibitem{huerta2023fair}
E.~Huerta, B.~Blaiszik, L.~C. Brinson, K.~E. Bouchard, D.~Diaz, C.~Doglioni, J.~M. Duarte, M.~Emani, I.~Foster, G.~Fox \emph{et~al.}, ``Fair for ai: An interdisciplinary and international community building perspective,'' \emph{Scientific data}, vol.~10, no.~1, p. 487, 2023.

\bibitem{ryu2022appfl}
M.~Ryu, Y.~Kim, K.~Kim, and R.~K. Madduri, ``{APPFL}: open-source software framework for privacy-preserving federated learning,'' in \emph{2022 IEEE International Parallel and Distributed Processing Symposium Workshops (IPDPSW)}.\hskip 1em plus 0.5em minus 0.4em\relax IEEE, 2022, pp. 1074--1083.

\bibitem{li2024advancesappflcomprehensiveextensible}
\BIBentryALTinterwordspacing
Z.~Li, S.~He, Z.~Yang, M.~Ryu, K.~Kim, and R.~Madduri, ``Advances in appfl: A comprehensive and extensible federated learning framework,'' 2024. [Online]. Available: \url{https://arxiv.org/abs/2409.11585}
\BIBentrySTDinterwordspacing

\bibitem{dwork2006differential}
C.~Dwork, ``Differential privacy,'' in \emph{International colloquium on automata, languages, and programming}.\hskip 1em plus 0.5em minus 0.4em\relax Springer, 2006, pp. 1--12.

\bibitem{beutel2020flower}
D.~J. Beutel, T.~Topal, A.~Mathur, X.~Qiu, J.~Fernandez-Marques, Y.~Gao, L.~Sani, K.~H. Li, T.~Parcollet, P.~P.~B. de~Gusm{\~a}o \emph{et~al.}, ``Flower: A friendly federated learning research framework,'' \emph{arXiv preprint arXiv:2007.14390}, 2020.

\bibitem{roth2022nvidia}
H.~R. Roth, Y.~Cheng, Y.~Wen, I.~Yang, Z.~Xu, Y.-T. Hsieh, K.~Kersten, A.~Harouni, C.~Zhao, K.~Lu \emph{et~al.}, ``{NVIDIA FLARE}: Federated learning from simulation to real-world,'' \emph{arXiv preprint arXiv:2210.13291}, 2022.

\bibitem{li2023appflx}
Z.~Li, S.~He, P.~Chaturvedi, T.-H. Hoang, M.~Ryu, E.~Huerta, V.~Kindratenko, J.~Fuhrman, M.~Giger, R.~Chard \emph{et~al.}, ``Appflx: Providing privacy-preserving cross-silo federated learning as a service,'' in \emph{2023 IEEE 19th International Conference on e-Science (e-Science)}.\hskip 1em plus 0.5em minus 0.4em\relax IEEE, 2023, pp. 1--4.

\bibitem{tuecke2016globus}
S.~Tuecke, R.~Ananthakrishnan, K.~Chard, M.~Lidman, B.~McCollam, S.~Rosen, and I.~Foster, ``Globus auth: A research identity and access management platform,'' in \emph{2016 IEEE 12th International Conference on e-Science (e-Science)}.\hskip 1em plus 0.5em minus 0.4em\relax IEEE, 2016, pp. 203--212.

\bibitem{hiniduma2024ai}
K.~Hiniduma, S.~Byna, J.~L. Bez, and R.~Madduri, ``Ai data readiness inspector (aidrin) for quantitative assessment of data readiness for ai,'' in \emph{Proceedings of the 36th International Conference on Scientific and Statistical Database Management}, 2024, pp. 1--12.

\bibitem{hoang2023enabling}
T.-H. Hoang, J.~Fuhrman, R.~Madduri, M.~Li, P.~Chaturvedi, Z.~Li, K.~Kim, M.~Ryu, R.~Chard, E.~Huerta \emph{et~al.}, ``Enabling end-to-end secure federated learning in biomedical research on heterogeneous computing environments with appflx,'' \emph{arXiv preprint arXiv:2312.08701}, 2023.

\bibitem{stahlschmidt2022multimodal}
S.~R. Stahlschmidt, B.~Ulfenborg, and J.~Synnergren, ``Multimodal deep learning for biomedical data fusion: a review,'' \emph{Briefings in Bioinformatics}, vol.~23, no.~2, p. bbab569, 2022.

\bibitem{tu2024towards}
T.~Tu, S.~Azizi, D.~Driess, M.~Schaekermann, M.~Amin, P.-C. Chang, A.~Carroll, C.~Lau, R.~Tanno, I.~Ktena \emph{et~al.}, ``Towards generalist biomedical ai,'' \emph{NEJM AI}, vol.~1, no.~3, p. AIoa2300138, 2024.

\bibitem{acosta2022multimodal}
J.~N. Acosta, G.~J. Falcone, P.~Rajpurkar, and E.~J. Topol, ``Multimodal biomedical ai,'' \emph{Nature Medicine}, vol.~28, no.~9, pp. 1773--1784, 2022.

\bibitem{fl_cancer}
S.~Pati, U.~Baid, B.~Edwards, M.~Sheller, S.-H. Wang, G.~A. Reina, P.~Foley, A.~Gruzdev, D.~Karkada, C.~Davatzikos \emph{et~al.}, ``Federated learning enables big data for rare cancer boundary detection,'' \emph{Nature Communications}, vol.~13, no.~1, p. 7346, 2022.

\bibitem{fl_medimg}
G.~Kaissis, A.~Ziller, J.~Passerat-Palmbach, T.~Ryffel, D.~Usynin, A.~Trask, I.~Lima~Jr, J.~Mancuso, F.~Jungmann, M.-M. Steinborn \emph{et~al.}, ``End-to-end privacy preserving deep learning on multi-institutional medical imaging,'' \emph{Nature Machine Intelligence}, vol.~3, no.~6, pp. 473--484, 2021.

\bibitem{ogier2022flamby}
J.~Ogier~du Terrail, S.-S. Ayed, E.~Cyffers, F.~Grimberg, C.~He, R.~Loeb, P.~Mangold, T.~Marchand, O.~Marfoq, E.~Mushtaq \emph{et~al.}, ``{FL}amby: Datasets and benchmarks for cross-silo federated learning in realistic healthcare settings,'' \emph{Advances in Neural Information Processing Systems}, vol.~35, pp. 5315--5334, 2022.

\bibitem{steyaert2023multimodal}
S.~Steyaert, M.~Pizurica, D.~Nagaraj, P.~Khandelwal, T.~Hernandez-Boussard, A.~J. Gentles, and O.~Gevaert, ``Multimodal data fusion for cancer biomarker discovery with deep learning,'' \emph{Nature machine intelligence}, vol.~5, no.~4, pp. 351--362, 2023.

\bibitem{kang2023benchmarking}
M.~Kang, H.~Song, S.~Park, D.~Yoo, and S.~Pereira, ``Benchmarking self-supervised learning on diverse pathology datasets,'' in \emph{Proceedings of the IEEE/CVF Conference on Computer Vision and Pattern Recognition}, 2023, pp. 3344--3354.

\bibitem{carrillo2024generation}
F.~Carrillo-Perez, M.~Pizurica, Y.~Zheng, T.~N. Nandi, R.~Madduri, J.~Shen, and O.~Gevaert, ``Generation of synthetic whole-slide image tiles of tumours from rna-sequencing data via cascaded diffusion models,'' \emph{Nature Biomedical Engineering}, pp. 1--13, 2024.

\bibitem{xiong2022unified}
B.~Xiong, X.~Yang, F.~Qi, and C.~Xu, ``A unified framework for multi-modal federated learning,'' \emph{Neurocomputing}, vol. 480, pp. 110--118, 2022.

\bibitem{che2023multimodal}
L.~Che, J.~Wang, Y.~Zhou, and F.~Ma, ``Multimodal federated learning: A survey,'' \emph{Sensors}, vol.~23, no.~15, p. 6986, 2023.

\bibitem{fu2021vf2boost}
F.~Fu, Y.~Shao, L.~Yu, J.~Jiang, H.~Xue, Y.~Tao, and B.~Cui, ``Vf2boost: Very fast vertical federated gradient boosting for cross-enterprise learning,'' in \emph{Proceedings of the 2021 International Conference on Management of Data}, 2021, pp. 563--576.

\bibitem{liu2024vertical}
Y.~Liu, Y.~Kang, T.~Zou, Y.~Pu, Y.~He, X.~Ye, Y.~Ouyang, Y.-Q. Zhang, and Q.~Yang, ``Vertical federated learning: Concepts, advances, and challenges,'' \emph{IEEE Transactions on Knowledge and Data Engineering}, 2024.

\bibitem{yu2023multimodal}
Q.~Yu, Y.~Liu, Y.~Wang, K.~Xu, and J.~Liu, ``Multimodal federated learning via contrastive representation ensemble,'' \emph{arXiv preprint arXiv:2302.08888}, 2023.

\bibitem{hatamizadeh2023gradient}
A.~Hatamizadeh, H.~Yin, P.~Molchanov, A.~Myronenko, W.~Li, P.~Dogra, A.~Feng, M.~G. Flores, J.~Kautz, D.~Xu \emph{et~al.}, ``Do gradient inversion attacks make federated learning unsafe?'' \emph{IEEE Transactions on Medical Imaging}, vol.~42, no.~7, pp. 2044--2056, 2023.

\bibitem{abad2020hierarchical}
M.~S.~H. Abad, E.~Ozfatura, D.~Gunduz, and O.~Ercetin, ``Hierarchical federated learning across heterogeneous cellular networks,'' in \emph{ICASSP 2020-2020 IEEE International Conference on Acoustics, Speech and Signal Processing (ICASSP)}.\hskip 1em plus 0.5em minus 0.4em\relax IEEE, 2020, pp. 8866--8870.

\bibitem{briggs2020federated}
C.~Briggs, Z.~Fan, and P.~Andras, ``Federated learning with hierarchical clustering of local updates to improve training on non-iid data,'' in \emph{2020 international joint conference on neural networks (IJCNN)}.\hskip 1em plus 0.5em minus 0.4em\relax IEEE, 2020, pp. 1--9.

\bibitem{thrun1995lifelong}
S.~Thrun and T.~M. Mitchell, ``Lifelong robot learning,'' \emph{Robotics and autonomous systems}, vol.~15, no. 1-2, pp. 25--46, 1995.

\bibitem{parisi2019continual}
G.~I. Parisi, R.~Kemker, J.~L. Part, C.~Kanan, and S.~Wermter, ``Continual lifelong learning with neural networks: A review,'' \emph{Neural networks}, vol. 113, pp. 54--71, 2019.

\bibitem{lee2020clinical}
C.~S. Lee and A.~Y. Lee, ``Clinical applications of continual learning machine learning,'' \emph{The Lancet Digital Health}, vol.~2, no.~6, pp. e279--e281, 2020.

\bibitem{li2023fedcompass}
Z.~Li, P.~Chaturvedi, S.~He, H.~Chen, G.~Singh, V.~Kindratenko, E.~A. Huerta, K.~Kim, and R.~Madduri, ``Fedcompass: efficient cross-silo federated learning on heterogeneous client devices using a computing power aware scheduler,'' \emph{arXiv preprint arXiv:2309.14675}, 2023.

\end{thebibliography}

\end{document}